\documentclass[runningheads]{llncs}
\usepackage[T1]{fontenc}
\usepackage[table]{xcolor} 
\usepackage{xcolor}
\usepackage{todonotes}

\usepackage{hyperref}
\usepackage{graphicx}
\begin{document}
\title{First Workshop on Building Innovative Research Systems for Digital Libraries (BIRDS 2025)}

\titlerunning{Building Innovative Research Systems for Digital Libraries}
%
\author{Christin Katharina Kreutz\inst{1,2}\orcidID{0000-0002-5075-7699} \and
Hermann Kroll\inst{3}\orcidID{0000-0001-9887-9276}}
\authorrunning{C. K. Kreutz and H. Kroll}
%
\institute{TH Mittelhessen - University of Applied Sciences, Gießen, Germany \and
Herder Institute, Marburg, Germany\\
\email{ckreutz@acm.org}\\
\and
TU Braunschweig, Braunschweig, Germany\\
\email{krollh@acm.org}}
\maketitle              
\begin{abstract}
We propose the first workshop on \textbf{B}uilding \textbf{I}nnovative \textbf{R}e\-search Systems for \textbf{D}igital Librarie\textbf{s} (BIRDS) to take place at TPDL 2025 as a full-day workshop.
BIRDS addresses practitioners working in digital libraries and GLAMs as well as researchers from computational domains such as data science, information retrieval, natural language processing, and data modelling. Our interdisciplinary workshop focuses on connecting members of both worlds. 
One of today's biggest challenges is the increasing information flood.
Large language models like ChatGPT seem to offer good performance for answering questions on the web.
So, shall we just build upon that idea and use chatbots in digital libraries?
Or do we need to design and develop specialized and effective access paths?
Answering these questions requires to connect different communities, practitioners from real digital libraries and researchers in the area of computer science.
In brief, our workshop's goal is thus to support researchers and practitioners to build the next generation of innovative and effective digital library systems.

\end{abstract}

\section{Background and Motivation}

Digital library professionals and representatives of GLAM institutes are domain experts. They typically have deep knowledge on their collections, metadata, users and their needs. Oftentimes they lack technical expertise or resources to implement practical solutions making use of current advancements from the computational domains such as machine learning, natural language processing, large language models or retrieval-augmented generation.
Conversely, researchers from computer science or similar domains usually have the corresponding computational expertise but are rarely aware of real-world challenges that exist in digital libraries. For instance, users of different domains may have completely different requirements for their access paths; see e.g. \cite{DBLP:conf/jcdl/KreutzBSSW23,DBLP:journals/corr/abs-2410-22358,DBLP:conf/jcdl/KrollKSB23,DBLP:journals/jodl/KrollPKKRB24}. 
In addition, fulling the user's needs asks for developing novel and effective access paths that have to meet constraints from libraries, e.g., having limited computational resources available or the prohibition to send material to ChatGPT. 
The goal of our workshop is thus to bridge these two communities, fostering collaboration leading to innovative practical applications in digital libraries.

Individual endeavours combining deep domain knowledge with computational expertise have proven to be fruitful, see for example the classification and automatic summarisation of historical newspaper articles in a collection of the national library of the Netherlands~\cite{DBLP:conf/ercimdl/KrollKCTB23}, the incorporation of multilingual retrieval in Europeana~\cite{10.1007/978-3-031-16802-4_8} or the improved access to datasets through automated data cleaning in HathiTrust~\cite{DBLP:conf/jcdl/HuJUD20}. 
Research like connecting the dots between scientific papers~\cite{metro_maps,linking_thoughts_within_scientific_papers}, infrastructures or literature graphs for scholarly content management~\cite{semantic_scholar,grapal,mag,openalex,infrastructures_for_scholarly_content}, and semantic enrichment of texts in the biomedical domain~\cite{biomed1,biomed2} show what is possible in information systems today. 
With our workshop we want to explicitly encourage and spark these serendipitous connections between disciplines.

Our BIRDS workshop loosely follows the concept of the successful Collab-a-thon at ECIR 2024~\cite{DBLP:journals/sigir/MacAvaneyRLPEKMFYSFSAFB24} by providing participants of the main conference an additional space to exchange or develop ideas surrounding practical digital library research and encouraging collaboration opportunities.  
That is why we explicitly do not ask for paper submission, but focus on scientific speed-dating, a keynote, panel, and group discussions with participants from both target groups.

\section{Goals and Expected Outcome}

With our BIRDS workshop we want to strengthen the collaboration in the community, connect domain experts with researchers in different contexts and academic states, and promote the construction of innovative research systems for digital libraries. Our goals are to:

\begin{itemize}
    \item Start a workshop series providing a forum to come up with collaborations between domain experts and researchers in digital libraries;
    \item Identify common challenges faced by digital libraries and explore how computational approaches could help address them;
    \item Provide a forum at TPDL to explicitly discuss rough ideas, problems and ongoing work;
    \item Produce a SIGIR Forum report describing the discussions and outcomes of our workshop;
    \item Continue this workshop series at subsequent Digital Library conferences.
\end{itemize}

Through interdisciplinary dialogue, we strive to advance applied research in digital libraries with the goal of pathing the way for better digital library systems.

\section{Format and Structure}
We anticipate a \textit{highly interactive}, \textit{in-person}, \textit{interdisciplinary} and \textit{engaging} \textit{full-day} workshop supporting collaboration and strengthening the community.
We aim to give room to participants to discuss existing libraries, potential practical problems, user requirements, and wishes. We strive to connect people with ideas or unsolved problems with people who are excited to help with possible solutions.
Therefore, a keynote, scientific speed-dating, a panel discussion and guided breakout sessions are planned. Table \ref{tab1} gives our tentative schedule.

\begin{table}[t]
\caption{Tentative schedule for the workshop.}\label{tab1}
\centering
\begin{tabular}{p{2.5cm}l}
\hline
Time & Agenda \\
\hline
9:00 -- 9:30 & Welcome and short introductions of all attendees\\
9:30 -- 10:30 & Keynote\\
\rowcolor{gray!20} 10:30 -- 11:00 & \textit{Coffee break} \\
11:00 -- 12:30 & Scientific speed-dating \\
\rowcolor{gray!20} 12:30 -- 13:30 & \textit{Lunch break}\\
13:30 -- 14:30 & Panel discussion \\
15:00 -- 15:30 & Breakout group discussions\\
\rowcolor{gray!20} 15:30 -- 16:00 & \textit{Coffee break}\\
16:00 -- 17:00 & Breakout group discussions\\
17:00 -- 17:30 & Summary of discussions and closing \\
\hline
\end{tabular}
\end{table}

\textbf{Welcome.}
In the first session of the day we will welcome all attendees and introduce the workshop. Afterwards, we will have a round of brief introductions for all attendees where they can state their background and a question regarding the workshop's topic they would want to have answered during the day.

\textbf{Keynote.}
Up next, a keynote will showcase a successful collaboration between practitioners providing the application scenario and posing the problems to be solved in the form of a digital library and researchers helping to overcome the shortcomings. We strive to invite a recognised researcher presenting an academic as well as a practitioner viewpoint.  

\textbf{Scientific Speed-Dating.}
After the coffee break we will have a scientific speed dating with all attendees. Attendees are paired together and get to know each other scientifically. After a short time (5-10 minutes), pairs are switched.

\textbf{Panel Discussion.}
After our lunch break, we will moderate a panel discussion with 5-7 invited panellists. The panellists will be from a diverse set of backgrounds with different seniority representing practical library as well as research perspectives. We intend to also invite one of the program chairs of this year's TPDL to potentially get some insights into the reviewing process and its current fit concerning applied digital library research. 

\textbf{Breakout Group Discussions.}
As the last bigger program point of our workshop day we will split participants into smaller groups for more focused discussions. These groups can discuss pre-defined topics or form organically, e.g., by common found interest through the scientific speed-dating. As for pre-defined topics we will provide prompts and questions such as \textit{LLMs and how to (not) use them} with questions on best practices and cases where LLMs do not work in practice, \textit{Working with OCR} with questions regarding current best practices and problems which are still faced in libraries or \textit{Evaluation of interfaces} with questions regarding best practices on questionnaire design and recruiting participants. Or more generally, questions around \textbf{novel and effective access paths} to digital libraries.

\textbf{Closing.}
In the last session of the day, the participants of the different breakout groups are asked to give short summaries on their discussions. There will be some time to further discuss the topics of the day in the plenum. Lastly, we invite participants to join us as co-authors in a SIGIR Forum report on the workshop before our workshop is closed.

\section{Intended Target Audience}
We invite participants from all backgrounds. The main target audience for our workshop are practitioners and researchers presenting their work in the main conference interested in building research systems in and for digital libraries. We strive to especially invite new members of the community, such as students or representatives of libraries, to take part in our workshop.
We expect to attract around 25 participants.

\section{Organisers}

\href{https://kreutzch.github.io}{\textbf{Christin Katharina Kreutz}} is a Tandem Professor for Data Science in the Humanities working as a practitioner at a GLAM institute (Herder Institute) and in academia (TH Mittelhessen~-- University of Applied Sciences).
Recently, she co-organised the Sim4IA workshop at SIGIR 2024, the SCOLIA workshop at ECIR 2025 and was poster chair for JCDL 2024. Her general research interests are information access systems and user behaviour/simulation. 

\href{https://www.hkroll.de}{\textbf{Hermann Kroll}} is a PostDoc at the Technical University of Braunschweig. His research focuses on effective access paths in digital libraries. Recently, he proposed a new retrieval paradigm, Narrative Information Access, for digital libraries and developed a real-world discovery system for the pharmaceutical domain. Beyond that, he published in digital libraries conferences such as JCDL, TPDL and ICADL and worked together with digital libraries, such as University Library of Braunschweig, the National Library of the Netherlands, the ZB MED in Germany and the University Library J. C. Senckenberg in Frankfurt am Main.

%
%
\bibliographystyle{splncs04}
\bibliography{bibliography}

\end{document}